\documentclass{sig-alternate-10pt}
\usepackage[margin=1in]{geometry}
\usepackage[sort&compress,numbers]{natbib} 

\usepackage{amssymb,url,color,graphicx,verbatim,hyperref,balance}
\newcommand{\bi}{\begin{itemize}}
\newcommand{\ei}{\end{itemize}}
\newcommand{\be}{\begin{enumerate}}
\newcommand{\ee}{\end{enumerate}}
\usepackage[normalem]{ulem}
\def\ignore#1{}
\newcommand{\shortversion}[1]{}
\newcommand{\daniel}[1]{\textcolor{red}{\textbf{Daniel}: #1}}
\newcommand{\antonio}[1]{\textcolor{blue}{\textbf{Antonio}: #1}}
\newtheorem{example}{Example}

\newtheorem{defin}{Definition}[section]

\newcommand{\bd}{\begin{defin}}
\newcommand{\ed}{\end{defin}}
\newcommand{\bex}{\begin{example}}
\newcommand{\eex}{\end{example}}

\begin{document}
\pagestyle{empty}


\title{A Call to Arms: Revisiting Database Design}

\numberofauthors{2}
\author{
\alignauthor
Antonio Badia\\
       \affaddr{University of Louisville, USA }\\
       \email{\small abadia@louisville.edu}
\alignauthor
Daniel Lemire\\
       \affaddr{Universit\'e du Qu\'ebec \`a Montr\'eal, Canada}\\
       \email{\small lemire@acm.org}
}
\date{}
\maketitle






\section{Introduction}
Good database design is crucial to obtain a sound, consistent 
database, and --- in turn --- good database design methodologies are
the best way to achieve the right design. 
These methodologies are taught to most Computer Science
undergraduates, as part of any Introduction to Database
class~\cite{impagliazzo2006computing}. They can be considered part of
the ``canon'', and indeed, the overall approach to database design has
been unchanged for years. Moreover, none of the major
database research assessments identify database design as a strategic
research
direction~\cite{Abiteboul:2005:LDR:1060710.1060718,Agrawal:2008:CRD:1462571.1462573,Bernstein:1998:ARD:306101.306137}.   

Should we conclude that database design is  a solved problem?
Our thesis is that {\em database design remains a critical unsolved
  problem}. Hence, it should be the subject of more research. 
 Our starting point is the observation that traditional database
 design is not used in practice --- and if it were used it would
 result in designs that are not well adapted to current
 environments~\cite{Avison:2003:DM:602421.602423}. In short,
   database design has failed to keep up with the times. 
  In this paper,  we put forth arguments to support our
    viewpoint, analyze the root causes of this situation and 
  suggest some avenues of research. 
 The point of view espoused here has been put forth more
or less explicitly in other places (see
\cite{Cohen:2009:MSN:1687553.1687576} for a recent and notable
example); but here we put together several strands that have received
isolated attention, and focus them on an issue that we feel is
particularly important --- database design.

In the next section (\S~\ref{problems}), we sketch the traditional
database design process: we argue that it manages to be, at the same
time, {\em over-engineered} and {\em under-engineered}. The
contradiction is only apparent: as any complex problem, this one is
multi-faceted. Traditional design does too little with respect to some
areas and too much with respect to others. 
In \S~\ref{roots}, we 
analyze the causes of the problems
presented in \S~\ref{problems}. 
We then briefly  the current status of research on database design (\S~\ref{sec:whatnow}).
Finally, we present some ideas for a research renewal in
\S~\ref{suggestions}. 

\section{Traditional  Modeling} 
\label{problems}
Relational modeling is usually broken down into three
steps: 
\bi 
\item \textbf{Conceptual modeling}, which includes {\em requirement
  gathering and specification}, and results in a {\em conceptual
  model} of the database. At this stage, the designer 
  focuses on issues of scope --- \emph{what belongs in the database?} --- and
  organization --- \emph{how is the information to be
  structured?} 
  Entity-relationship
  diagrams~\cite{chen1976entity} and UML class models are the two
  best known conceptual models, but not the only ones; alternatives
  like Object Role Modeling have been proposed~\cite{HAL}. 

\item \textbf{Logical modeling}, which takes as input the conceptual
  model produced in the previous step and yields a database
  schema. This step is well  developed~\cite{TLNJ}. Normalization
  enforces
  {\em functional dependencies} by removing redundancy.

\item \textbf{Physical modeling}, which takes as input the database
  schema produced in the previous step and produces storage structures
  to  implement the schema in computer systems. 
  It can be automated to a large
  extent~\cite{CHAU,Gebaly:2008:RAP:1353343.1353365}.
\ei

Each step focuses on only one aspect of the problem which helps tame the complexity. Also, each step produces an output
that feeds into the next step, creating a  linear structure
that is easy to follow.

\subsection{Problems with the traditional approach}
The problem of database design is difficult, and it
encompasses issues that may not be amenable to 
formalization~\cite{Roddick:2007:PSD:1793834.1793852}. Hence, any method is likely to have some limitations and
drawbacks. However, this is not a reason to ignore the serious
problems that the traditional approach is running into. Here we
summarize what we see, from our experience and perspective, as the
most troublesome issues. 

\subsubsection{Failures of use and guidance}

We claim that the traditional approach is not followed in practice.
Indeed, Fitzgerald et al.~\cite{fitzgerald1999information} found that only about 11\% of
the consulted organizations claimed using an unmodified commercial
information system methodology. 
Furthermore, 
Brodie and Liu~\cite{brodie2010} report that while 90\% of
all information systems inside a Fortune~100 company are relational,
they could not find a single instance of an entity-relationship modeling
in over ten~such large corporations. The lack of modeling is not due to the
lack of complexity: they report that a typical Fortune~100 company
has about 10~thousand different information systems, that a typical relational
database is made of over 100~tables, each containing between 50 to
200~attributes.  
  Formalized conceptual models, as well as the theory
  developed around normalization, are not used. Physical modeling is frequently delayed 
  until performance problems arise.
 In a very real way, we have entered a post-methodological era as far as the
 design of information systems is concerned~\cite{Avison:2003:DM:602421.602423}.
 The emergence of the Web has coincided with the death of the dominant
 methods based on the analytic thought and lead to the emergence of
 sense-making as a primary paradigm.

If one agrees that the traditional method is not used, the obvious
question  is: Why? Why do practitioners dismiss a method that
has a solid theoretical basis and is the distillation of years of
thought? 
 It would be easy (and tempting!) to blame the designers or their
 training. 
  But the tools themselves share a good part of the blame. They fail to give what designers need
  most, {\em guidance as to how to apply them}:
for conceptual models, not enough guidance is given as how to create
one, how to assess its quality,  and --- importantly --- how to
handle all information that does not fit into the conceptual model but
may be relevant later for data quality of other purposes.

A critical failure in the traditional approach is that there is little
guidance on how to discover important information (e.g., functional dependencies) 
in the real world. It would not be a concern if the rest of the design
assumed that we lacked information.
Yet unless we have  {\em all} functional dependencies, there is no
guarantee of normal form in the logical design.  Thus,
the logical design phase is \emph{brittle}.  

Ironically, the step where most research has focused on giving
guidance is the last one, physical design~\cite{CHAU}, perhaps because
it is easier to simulate realistically the problems and their solutions in a laboratory.
However, this
third step relies on the previous ones; while it can sometimes result in
modifications of the database schema --- as when denormalization is
recommended, most approaches still assume that a schema has been well
designed.
By analogy, we could say that we know how to build the walls, 
as long as the foundation of the house is, somehow, done properly.


\subsubsection{Failures of imagination}
\label{sec:failuresofimagination}

Even if one were to follow the steps of the traditional design method,
and have a perfectly normalized, by-the-book database, what does one
obtain? 

We consider database design a matter of semantics: we are trying to
capture the semantics of a domain, to represent information about that
domain faithfully, and to (only) allow operations with the data that are
meaningful. But traditional database design focuses on structure. In
exchange for all the effort, we have insufficient semantics. This is the sense in which databases are {\em under-engineered}. 

Consider, for instance, the problem of {\em information
    integration}~\cite{II3,II1,II2}. 
Relational databases fail to provide enough information 
    to determine automatically whether
two databases contain information about distinct, overlapping, or
similar domains. And yet, integration of information is increasingly critical:
40\% of the cost associated with information systems is due to data
integration problems~\cite{brodie2010}.  
To exemplify this trend toward greater integration and collaboration
even in the most conservative settings, consider that 
the 9/11~Commission report urged the intelligence community to move from its
 need-to-know standard to a need-to-share
 approach~\cite{jones2007}. Experts believe 
that the 9/11 tragedy could have been avoided with better data integration.
The traditional way to design databases does not capture enough
information to enable information integration ---  in fact, it
falls short of capturing precisely the kind of information that would be
  more valuable for integration. Hence, traditional design not only
fails to alleviate the problem, {\em it is helping to perpetuate it}.
Most data integration approaches start by trying to determine the
similarity between attributes.\footnote{Several 
  approaches rely on statistical properties of data, and choose not to
try to interpret it~\cite{KN}. It is unclear whether this is done in
search of generality or due to need; but we believe that, while this
approach provides important information, statistical properties cannot
{\em establish} semantic similarity {\em by themselves}---but see
Halevy et al.~\cite{Halevy} for a different viewpoint.} Since most
design approaches treat attributes as barely more than labels, one has
usually only a string to work with: information {\em about} attributes
(metadata) is usually absent~\cite{SSR2}. As long as design focuses on
how to structure attributes in tables and not in what attributes mean,
the problem will be with us. 
In the end, we ask practitioners to follow a model that is
  demanding and  yields, in return, some very limited results.

\ignore{
What do we mean by a focus on structure? 
 Assume attributes $A$, $B$, $C$, completely
unrelated. Then the table $(A,B,C)$ has only the trivial functional
dependency :$ABC \rightarrow ABC$, and is in BCNF.  But this does not
help. Thus, it is clear that a perfectly normalized database may not
make intuitive sense.
Second, other normal forms (like domain-join normal form) are not
algorithmic and may not be doable.
Third, lack of information on attributes and other information not present
in conceptual models leads to lack of checks, assertions
and triggers.
}
The lack of appropriate metadata is even more acute in 
new  applications, ranging from financial to legal systems. A
prominent example is {\em e-science}:  scientists need not only to
  store larger and larger amounts of data. They also need to be able
  to assess provenance~\cite{Simmhan:2005:SDP:1084805.1084812},
 access rights,  workflows, etc. in order to comply
with ever increasing regulations, to be able to share the data, and
to achieve the goal of {\em reproducible
research}~\cite{Stolteetal}. On this, traditional 
  design offers no guidance.

To make matters worse, the focus on structure creates
rigidity. Kiely and Fitzgerald~\cite{kiely2005investigation} found
that traditional information systems development methods were sometimes
perceived to be  of limited use within modern projects because
they are too cumbersome and inflexible.
This is the sense in which databases may be considered {\em
  over-engineered}.
Consider the NoSQL movement~\cite{leavitt2010will}. A large force
behind it are programmers  
for which database design makes no sense. Tired of the rigid structure
of relational databases, other systems (Raven DB,\footnote{{\tt
    http://ravendb.net}} Amazon's 
SimpleDB,\footnote{{\tt http://aws.amazon.com/simpledb/ }} Apache's
CouchDB,\footnote{{\tt http://couchdb.apache.org/ }}
MongoDB\footnote{{\tt http://www.mongodb.org/ }}) are emerging.  
What good is it to design if it fails to make the developers
more productive? 
Unfortunately, the
mismatch between objects and program structures on one hand, and database
structures on the other, is still largely unresolved. Motivated by
this problem, Microsoft has proposed the Language-Integrated Query (LINQ)
framework~\cite{Meijer:2006:LRO:1142473.1142552}. Other initiatives to
bridge the gap have been developed over the years --- witness to the fact
that the problem is still with us.


\section{Why does it fail?}
\label{roots}
 The traditional design method was developed in the early seventies, when
 mainframes dominated information technology. It is in this era that the
relational~\cite{Codd:1970:RMD:362384.362685} and  entity-relationship
models~\cite{chen1976entity}  were invented.
 Accordingly,
there are several assumptions behind the traditional design which reflect
its age:

\noindent $\bullet$ \emph{Users are faceless objects for whom (or on
  whose behalf) 
  the systems are designed~\cite{ISJ:ISJ336}}.\footnote{In the quote
  sometimes attributed to Frederick the Great, 
  it's ``everything for the people, 
  but without the people''. 
  }  In the seventies, the management
  of data  was left in the hands of few experts who served 
the needs of technologically unsophisticated employees.  Nowadays, 
 the boundaries between users, whether they are employees
or clients, and developers are blurred~\cite{millerand2010users}. This
is best illustrated with 
how \emph{hashtags} emerged on the microblogging  platform Twitter. Hashtags
are a metadata convention among Twitter users~\cite{Kwak:2010:TSN:1772690.1772751}, in the spirit of
folksonomies~\cite{Siersdorfer:2009:SRS:1557914.1557959}. Yet Twitter
itself had no support for metadata. We can trace back the current
convention to a 
single user 
 who informally proposed it in a 140-character
post in August 2007. Later, Twitter engineers recognized the convention
and added software support for it. For example, Twitter detects ``trending
topics'' using popular hashtags. A few other conventions, like the ``retweet''
were first initiated by the users.
Sundara Nagarajan has recently expressed the same idea~\cite{SUN}:
``Empowered end users cause application systems to evolve at tremendous
speeds and continuously create new requirements for
interoperation. For instance, a social networking site user can add
content and pointers from a website, by simply dragging and
dropping. The evolution of mashups that combine data and functionality
from multiple sources is another example of this new design
paradigm. This is leading to the evolution of the user experience,
along with computation and data management.''
When systems are designed without the users, a lack
of user engagement may result: 93\% of all accounts in Business
Intelligence systems are never
used~\cite{Meredith:2010:FMS:1860745.1860761}.  

\noindent $\bullet$ \emph{The information system is strongly consistent.} 
It has been estimated that Google alone has more than 1~million
 servers.   Using
 cloud computing, anyone can use a distributed network of servers at a
 modest cost. 
With multiply located servers and deeply integrated web services,
 the CAP theorem~\cite{gilbert2002brewer} implies
in practice that we have to choose between strong consistency and
strong availability: we cannot have both.
As a possible illustration of this constraint, the recent
failure of an Oracle database at JPMorgan Chase, which froze
\$132 million in assets and lost thousands of loan applications,
was blamed on an  database~\cite{monashjpmorgan} which
required strong consistency for all data. 

\noindent $\bullet$ \emph{Semantics is absolute.}
The original design assumed a centralized
architecture. This architectural assumption had a reflection on the
conceptual level, where 
one main viewpoint was assumed. While semantic
relativism is pointed out~\cite{EN,RG,SKS},  a
choice must be made for a single model: there is no mechanisms to derive other. 
Yet when different systems must interact routinely, we cannot expect
that they all share the same viewpoint.

\noindent $\bullet$ \emph{The models are static.} In the traditional
setting, there is little need for evolution. 
Yet  databases, even in large conventional corporations, are fast
evolving: 30\% of all information 
systems are modified significantly every year~\cite{brodie2010} in Fortune~100 companies. 
Chen, the father of the entity-relationship model, recently
recognized the  difficulty by pointing out the inability of existing
modeling techniques to cope with fast-varying world
states~\cite{springerlink:10.1007/11901181-1}. 

\section{What now?}
\label{sec:whatnow}


Despite these  difficulties, {\em research on database design 
has failed to make major progress in the last ten years or so}.
This is not to say that no
research is done.
For instance, the series of conferences on entity-relationship
modeling~\cite{ER}, while not totally focused on design
issues, devote most of the program  
 to them.
Theoretical work is still ongoing in logical
modeling~\cite{Kolahi:2008:IAW:1670243.1670248}. 
Physical design research is still strong, sometimes driven by
database vendors~\cite{Agrawal:2005:DTA:1066157.1066292}. 
And there is still a community of dedicated researchers including
notable researchers like  Thalheim{~\cite{Thal}} and 
Oliv\'e~\cite{Oli} among others.

But overall  the topic is not widely pursued. For instance, if one
checks the last 10 years of the ``major''  database
conferences (SIGMOD, VLDB, ICDE), the number of papers in 
database design is  low: leaving aside {\em physical design} 
(that is, the fine tuning of storage structures for better
performance), we found less than ten  talks 
{\em mainly} about database
design~\cite{Andritsos,Beyeretal,Golabetal,Jarkeetal,LuoWu, 
  Maddenetal,Stolteetal,Tunkelang}
 and most of them examine design issues within the confines of a restricted
 context
(sensor databases~\cite{LuoWu,
  Maddenetal}, XML schema 
evolution~\cite{Beyeretal}, user interaction~\cite{Tunkelang}, scientific databases~\cite{Stolteetal}, data
warehousing~\cite{Jarkeetal}).\footnote{
	We do not claim any
  statistical validity for this observation. For one, the sample used is
  limited --- more conferences, and certainly some journals, should be
  included. For another, there is a
  subjective aspect to this analysis. For the sake of transparency, we
  explain our method: first, the web page of each conference,
  reached through the web site of the organization behind the
  conference  was used: this gives access to session titles,
  paper titles and sometimes abstracts. A search was made for
  keyword ``design'', another for keyword ``normal'' (to obtain
 'normalization' and so on), and another for ``semantic''. We
  checked the title and abstract for each match (the last
  keyword generated quite a few matches). As stated, papers on
 physical design (database tuning and index design) were
  excluded. The list of papers obtained is given for interested 
  readers to judge by themselves in the references above. 
 Others may reasonably disagree exactly as to what to include,
  i.e., how to define 'mainly about database design' (but
  note that we include an invited talk and two tutorials!).  However,
  unless one uses a generous notion of 'database design',
  we believe other people's results will be in the same order
  of magnitude as  ours.
}
There are certainly papers which, without having design as their main
goal, bring considerable contributions to the table. For instance, the
research on {\em dataspaces} by Halevy et
al.~\cite{Halevy:2006:PDS:1142351.1142352} has 
brought forth the possibility of databases where the schema is
implicit or at least not separated from the data and can evolve with
it. 
It opens up some 
possibilities, but 
no paper on this project is about design {\em per
  se}. Likewise, research on semistructured data (e.g., XML) has exposed the
database community to the thought that design must be more
flexible~\cite{Barbosa:2005:DIM:1083592.1083608}.
However, little of 
this seems to have percolated to more traditional (relational, object
relational)  data models, and the design methodology for them. 
Finally, the total number of contributions remain
low, even including this work, for such a crucial topic.

 The fact is, {\em traditional database design is not a mainstream research topic}.
 We believe that this is due to two main facts: first, for most
researchers, 
 work on database conceptual models is seen as too difficult,
because the subject is ``soft'', 
not clearly formalized, and does not
yield itself well to the typical paper that one expects to see
published in most conferences and technical journals. Second, 
work on relational design is considered useless as the topic is
commonly taken as basically a mature and closed one. Certainly, there
is always some more work that can be done (for instance, extending the
idea of key and functional dependence to other models, like XML, has
received some recent attention~\cite{Chen:2010:FDX:1927585.1927598},
as well as extending the concept itself to 'soft' functional
dependencies~\cite{Ilyas:2004:CAD:1007568.1007641}), 
but the subject is often considered  ``a solved problem''.

On teaching, we note that while some textbooks are quite
good about pointing out to students the limits and difficulties of the
process, others simply gloss over the issues and give the impression
that this a ``case closed'' situation --- which may contribute to the
lack of research in the area. 

\section{What next?}
\label{suggestions}
Traditional database design fails to provide the tools needed to
design databases in today's environment, but researchers have not
updated or expanded the methodologies enough to keep up with the times. 
Should we continue teaching methodologies which disappoint practitioners?

A first step towards renewed emphasis on database design research is
to come up with a fresh and timely approach. Different
researchers will likely have different viewpoints as to what are the
most crucial or interesting problems. We submit the following
research plan to open up a discussion.

\paragraph{Design for a distributed world}

\begin{itemize}

\item We must update database design methodologies for new
environments that did not exist in the 1970s. Though there were many
failed attempts to replace the ACID-compliant
relational database systems with \emph{better} alternatives, the landscape 
has finally begun changing with the adoption of cloud computing.
For example, the data  consistency  requirements (and other issues
affecting distribution) should be made explicit during the design
phase, so that they can be exploited when deciding an architecture.
In fact, many NoSQL designs assume that most operations can be kept
local in order to ensure scalability, which means that one needs to
know which data is likely to be involved in a transaction (logically
related) in order to distribute the data (\cite{Cattell2} makes the
same point, implicitly). Along the same lines, 
deciding what can be made eventually consistent (versus what needs to
be kept consistent at all times), and what to do in the face of
inconsistency, should be based on the semantics of data. Hence, such
issues should be part of the design phase.

\item    
  It is fashionable to talk about Big Data: one the main driver
  being this trend is our ability to quickly integrate diverse
  data sets to create new services. 
  Correspondingly, {\em easier data integration should become one of
    the primary goals of good database design}. 
  Another issue that Big Data brings is the distributed nature of the
  model. Do we need a 'distributed design' approach? For instance,
  should design produce more or less independent modules or 'chunks'
  of connected data, which can in turn be connected to each other in
  one or more ways? 
  How would such a distributed design relate to 
  Berners-Lee's linked data~\cite{berners2006linked}? 
   Or perhaps we should propose methodologies which,
 instead of starting from a clean slate, begin with the existing
 schemas (both within the organization, and public ones) and build on
 top of them. Should we shift the focus towards {\em extensions of what
   there is}? 

 \end{itemize}

\begin{samepage}
\paragraph{Rethink functional dependencies}\begin{itemize}
\item  
 If Helland is right and normalization is 
 for sissies (\cite{Helland}), then one should question the focus on 
 functional dependencies in database design. If this idea seems
 far-fetched, recall that data warehousing practitioners proposed a
  different design methodology (the star schema) that does
 not use the idea of functional dependency at all (rationalization of
 star schema in normal form came after the fact). 
The question then is whether 
 there are other concept or concepts that can replace 'functional
 dependence' and be a good basis for design.
\item 
 We know that enforcing functional dependencies in the schema
 is insufficient to ensure that the data is semantically consistent.
 There are many rules, some expressible as constraints, assertions, or
 triggers, that could be enforced  to ensure meaningful
 data. But current design mostly ignores this information.
 Shouldn't we attempt to capture this information, which is most likely
 to have an  impact on the quality of our data, during the design? (In
 which case we need to define a way to  measure the impact of
 different types of rules in data quality and consistency.) How can
 these various rules be used {\em together} in design? Note that to
 answer this question one has to answer other, more basic, questions:
 how do these different rules interact?

\item Much of the  database-design
courses focus on  functional dependence and normal form. 
It is often implied that the physical design ought to be 
a straight-forward application of the logical design.
This is because, once, the equation {\em
    one relation = one table = one file} held for virtually all
  relational systems.  Yet it not longer applies.
For example,   many distributed or column-oriented database systems
replicate data for speed or reliability.  Is it time to {\em
  completely} separate   logical design from physical 
  design, i.e. consider the relation as a purely conceptual entity?
 \end{itemize}
\end{samepage}

\paragraph{Design for imperfect knowledge}
 
 \begin{itemize}
 \item  We must cope with 
incomplete information (about the domain, the users, etc.) since in
real systems, the scope or boundary of a database, or its future
usage, is often uncertain~\cite{Magnani:2010:SUM:1805286.1805291}. Thus, design should proceed with as few
assumptions as possible. Until now, a certain {\em closed world
  assumption} mentality trickles all the way from  the conceptual
model to the database. Clearly, we live in an {\em open} world. 
Should be consider schemas as {\em descriptive} instead
of {\em prescriptive}, which is what they are now? If so, what to do
with data that does not follow the schema? Should any such data be
allowed?  Given the difficulty of determining in advance the type of
data that the system may have to deal with, should the design include,
for instance, a description of data that should {\em not} be allowed,
and leave the database open to all other data? 
To some extend, XML schema languages (e.g., XML~Schema and Relax~NG)
seem to  adopt such a permissive attitude, with the added requirement that the
data be structured in a hierarchical manner. Unfortunately, our experience
is that the process is burdensome and is not widely used~\cite{braynonewlanguage}. Are
there lightweight alternatives?

\item In turn, adopting an open world point of view will make it
  easier to support collaborative, evolutionary design as an
  integrated part of the
  workflow~\cite{Cohen:2009:MSN:1687553.1687576}. The issue here is, 
how do we design databases with  {\em open}-world model while insuring
the necessary consistency?  If we are going to give permission to
users to modify a schema, how much freedom should users have? For
instance, one could study {\em whether design can be crowdsourced}
(and if so, how and under what constraints).
In general, one need to decide what kind of changes can be support,
whether they come from the users or from a designer. A deeper study of
{\em database evolution} could be of help here: could a system be
designed that adapts its storage to changing schemas and requirements? 
Physical design is currently focused on {\em query workload}, that is,
it adapts itself to the (changing) requirements posed by the database
queries. Could some of these ideas be used to make the system reactive
to changes in the schema?  We find interesting that 
functional dependencies can be (roughly)  classified as {\em natural}
(one that reflects an invariant in the world: a person
has only one height) or {\em artificial} (one that reflects a
convention: each employee has to attend X meetings a month). The
former are quite stable, but the latter are subject to change (note
that all so-called {\em business rules} are artificial!). Should a
system be able to cope with changes in artificial dependencies (old
ones cease to hold, new ones are added)? 

\item 
  New data stores in the NoSQL
  movement use non-relational data models: 
  key/value, documents,  extensible records~\cite{Cattell}. Probably
  the first research task for such data models is a clarification of
  their exact structure and properties, since the terms are used
  somewhat loosely. But an immediate second is to decide whether they
  require a different approach to design (after all, even NoSQL data
  stores require design) or, to the contrary, whether design 
  decisions can be kept independent of the data model. The question is
  not as trivial as it may seem: some of these new models allow an
  {\em open schema}, that is, one where the user can add attributes at
  will, while others still require, like relational databases, a {\em
    closed schema}, that is, one where all possible attributes are
  declared beforehand --- yet others, like extensible records, combine
  both parts. 

\item 
  Though there has been much work done on probabilistic 
  databases~\cite{Dalvi:2009:PDD:1538788.1538810}  and soft functional
  dependencies~\cite{Ilyas:2004:CAD:1007568.1007641}, such subjects
  remain almost entirely distinct from database design.  Yet semantics
  are not always absolute: some relationships are merely  almost
  always true.  Thus, it is likely that there are many more soft
  dependencies or conditional dependencies than 'standard' functional
  dependencies. (A conditional dependency is one that holds only under
  certain circumstances. For example, at some places, a married couple
  is always made of  a man and a woman, but not at others.)
 Current design practices tend to ignore all  functional 
 dependencies but the standard ones, which are but an extreme
 case~\cite{10.1109/TKDE.2010.154}. 
 Should we make room in  database-design
 methodologies for probabilistic metadata and several types of
 dependencies?  If so, how would different types of dependencies be
 used? How would they behave when put together?

 \end{itemize}

No doubt, different researchers will have different viewpoints on
these issues. Some may object to some of the challenges included here;
others may wish to direct attention to other problems not included
here. We stress again that this plan is meant to start the discussion;
let the debate begin.


\bibliographystyle{abbrv} 
\balance\bibliography{sigmod.record.full}

\end{document}